\documentstyle[prl,aps,graphicx,epsf,amssymb]{revtex}

\newcommand{\be}{\begin{equation}}
\newcommand{\ee}{\end{equation}}
\newcommand{\bi}{\begin{itemize}}
\newcommand{\ei}{\end{itemize}}

\newcommand{\vxi}{{\mathbf \xi}}
\newcommand{\x}{{\mathbf x}}
\newcommand{\opx}{\widehat{{\mathbf x}}}
\newcommand{\X}{{\mathbf X}}
\newcommand{\Y}{{\mathbf Y}}

\newcommand{\y}{{\mathbf y}}

\newcommand{\J}{{\mathbf J}}

\newcommand{\A}{{\widehat{A}}}
\newcommand{\tA}{{\widetilde{A}}}

\newcommand{\opR}{{\widehat{R}}}
\newcommand{\opn}{{\widehat{n}}}

\newcommand{\opP}{{\widehat{P}}}
\newcommand{\opT}{{\widehat{T}}}
\newcommand{\oprho}{{\widehat{\rho}}}

\newcommand{\opH}{\widehat{H}}

\newcommand{\oq}{\widehat{q}}
\newcommand{\op}{\widehat{p}}

\newcommand{\zero}{{\mathbf 0}}

\newcommand{\tW}{\widetilde{W}}

\newcommand{\tr}{\mathop{Tr}}

\newcommand{\der}{\partial}

\begin{document}

\title{Parity Measurements, Decoherence and Spiky Wigner Functions}

\author{A. M. Ozorio de Almeida\footnote{ozorio@cbpf.br},
O. Brodier\footnote{brodier@cbpf.br}}

\address{Centro Brasileiro de Pesquisas Fisicas, 
Rua Xavier Sigaud 150, 22290-180, 
Rio de Janeiro, R.J., Brazil.}

\maketitle

\begin{abstract}

Notwithstanding radical conceptual differences between classical and quantum mechanics, it is usually assumed that physical measurements concern observables common to both theories . Not so with the eigenvalues ($\pm 1$) of the parity operator. The effect of such a measurement on a mixture of even and odd states of the harmonic oscillator is akin to separating at a single stroke a pair of shuffled card decks: the result is a set of definite parity, though otherwise mixed. The Wigner function should be a sensitive probe for this phenomenon, for it can be interpreted as the expectation value of the parity operator. We here derive the general form of Wigner functions $W_{\pm}$, resulting from an ideal parity measurement on $W(\x)$. Even if $W(\x)$ resembles a classical distribution, $W_{\pm}$ displays a quantum spike, which is positive for $W_+$ and negative for $W_-$. However we conjecture that $W_+$ always has negative values.

\end{abstract}

The parity operator, an observable currently measured in quantum optics
\cite{EngSte93}\cite{LutDav97}\cite{BerAuf02}, 
can be written $\opR_{\zero}=(-1)^{\opn}$. Here $\opn=\opH/\hbar$, 
where $\hbar$ is the Planck constant and $\opH$ 
is the Hamiltonian of the harmonic oscillator with unit frequency. Its 
expectation value is proportionnal to the Wigner function $W(\x)$\cite{Wig32}
\cite{HilOCo84}\cite{Sch:book} of the system, taken at the point $\x=\zero$.
One has indeed\cite{Roy77} $W(\zero)=\tr{\oprho}\opR_{\zero}/\pi\hbar$.
The direct measurement of the Wigner function results from counting the relative proportions of the eigenvalues $\pm 1$ for repeated parity measurements \cite{BerAuf02}.

The apparent classical nature of the Wigner function, which allows the calculation of quantum expectations as phase space integrals, is belied by narrow negative oscillations in most pure states. If the quantum system evolves in contact with the external environment, decoherence\cite{GiuJoo:book}\cite{Zur03} smoothes the Wigner function and eventually erases the negative fringes. The threshold time for complete positivity\cite{DioKie02}\cite{BroAlm03} of the Wigner function is independent of the initial pure state, within the Markovian approximation which treats the environment statistically (with no memory effects) if one further restricts the coupling to be linear and the internal Hamiltonian to be quadratic. Then the picture\cite{BroAlm03} is that of a real function transported by a linear coordinate transformation in phase space, coarse-grained by a widening Gaussian window due to the external coupling.  
A  measurement of the parity operator generates a central spike of maximum modulus\cite{EngMor02}, $W(\zero)=\pm(\pi\hbar)^{-1}$, on the previously smoothed Wigner function.  Furthermore, a weaker pattern of fringes reemerges, resembling those of pure states. Quantitative measures still indicate overall decoherence, confirmed by the coarse-graining of $W(\x) $, far from the reflection centre. Even so, the positivity threshold for the further Markovian evolution of an odd state, $W_{-}(\x) $, is the same as for a pure state, generally exceeding the time for ordinary mixed states to loose their negative regions. The sharp spike of $W_{\pm}(\x) $ signals the full recovery of quantum parity as a consequence of its experimental measurement.

The quantum operator, $\opR_{\x} $, corresponds to a (classical) reflection through the phase space point $\x=(p,q) $, i.e. other points $\x'\rightarrow 2\x-\x' $. It is possible to specify $\opR_{\x} $ by a superposition of projection operators\cite{Roy77},$|q\rangle\langle q'| $:
\be
\opR_{\x}=\frac{1}{2}\int~dq'~|q-\frac{q'}{2}\rangle ~e^{-i\frac{pq'}{\hbar}}
~\langle q+\frac{q'}{2}|.
\ee
 From this we obtain the well known definition of the Weyl symbol for an arbitrary operator, $\A $, as\cite{WeyRob:book}\cite{Gro76}\cite{Alm98}
\be
A(\x)=2\tr{ \opR_{\x}\A} = \int ~dq'~\langle q+\frac{q'}{2}|\A|q-\frac{q'}{2}
\rangle ~e^{-i\frac{pq'}{\hbar}}.
\label{defWeyl}
\ee
Here,``$\tr (ace)$'' denotes the sum over all eigenvalues of an operator.

In the case of the density operator, $\oprho $, it is conveniently normalized to obtain the Wigner function\cite{Roy77}, $W(\x)=\rho(\x)/2\pi\hbar $, which allows the computation of averages\cite{HilOCo84}\cite{Sch:book}\cite{Alm98}
\be
\langle\A\rangle = \tr{\oprho\A} = \int~d\x~ A(\x) W(\x),
\ee
as if $W(\x) $ were a classical probability density. However, the Wigner function is never concentrated on a single phase space point, so there is no contradiction with Heisenberg's uncertainty principle.

 The result of a parity measurement for reflection through a point $\X $ on a quantum system described by the density operator $\oprho $ must be one of the alternatives 
\be
\left\{ \begin{array}{ccc}
\oprho_{+}^{\X} & = & 
\frac{\opP_+^{\X}\oprho\opP_+^{\X}}{\tr{\oprho\opP_+^{\X}}} \cr
\oprho_{-}^{\X} & = & 
\frac{\opP_-^{\X}\oprho\opP_-^{\X}}{\tr{\oprho\opP_-^{\X}}}
\end{array}
\right. ,
\label{proj-dens}
\ee
allowed by standard quantum theory\cite{CohDiu:book}, where the orthogonal projection operators for each parity are\cite{Roy77}  
\be
\opP_{\pm}^{\X} = \frac{1}{2}\left(1\pm\pi\hbar W(\X) \right).
\label{proj-op}
\ee
An obvious procedure for calculating $ \oprho_{\pm}^{\X}$ is to use an orthogonal basis of even and odd states, such as harmonic oscillator eigenstates for $\opR_{\zero} $. However, such unwieldly calculations are conveniently short-circuited by the Wigner-Weyl representation which is already built upon reflection operators. Indeed, from (\ref{defWeyl}) and (\ref{proj-op}) we immediately obtain the denominator in (\ref{proj-dens}) as 
\be
\tr{\oprho\opP_{\pm}^{\X}} = \frac{1}{2}\left(1\pm\pi\hbar W(\x)\right).
\ee
The full Wigner function corresponding to $\oprho_{\pm}^{\X} $ depends on the symplectic matrix
\be
\J = \left( \begin{array}{cc} 
0 & -1 \cr
1 & 0 \end{array} \right)
\ee
and the Fourier transform
\be
\tW(\vxi)=\frac{1}{2\pi\hbar}\int~d\x ~W(\x) 
\exp{\left(\frac{i}{\hbar}\x\cdot\J\vxi\right)},
\ee
which is itself a bona fide representation of the density matrix, known as the chord function\cite{Alm98}, or the characteristic function in quantum optics. Then the Wigner functions corresponding to the projected densities (\ref{proj-dens}) are
\be
W_{\pm}^{\X}(\x) = \frac{1}{2}\frac{W(\x)+W(2\X-\x)\pm 
4\Re{ \tW\left(2(\x-\X)\right)\exp{\left(-\frac{2i}{\hbar}\x\cdot\J\X\right)} }
}{1\pm\pi\hbar W(\X)}
\label{Wig-proj-dens}
\ee
where $\Re $ denotes the real part of a number. This formula generalizes the specific formula for $W_{\pm}^{\zero}(\x) $ in the case of circular symmetry\cite{EngSte93}. 

The derivation of (\ref{Wig-proj-dens}) is straightforward if one combines
the reflection operators $\opR_{\x} $ with the translation operators,
\be
\opT_{\vxi} = \exp{\left(\frac{i}{\hbar}\opx\cdot\J\vxi\right)}=
\exp{\left(\frac{i}{\hbar}\left(\vxi_p\oq-\vxi_q\op\right)\right)},
\ee
to form a quantum version\cite{Alm98} of the affine group of the translations and reflections\cite{Cox:book} in phase space:
\be
\begin{array}{ccccccc}
\opT_{\vxi_1}\opT_{\vxi_2} & = & \opT_{\vxi_1+\vxi_2}
\exp{\left(-\frac{i}{2\hbar}\vxi_1\cdot\J\vxi_2\right)} & ; &
\opT_{\vxi}\opR_{\x} & = & \opR_{\x+\frac{\vxi}{2}}
\exp{\left(-\frac{i}{\hbar}\x\cdot\J\vxi\right)}; \cr
\opR_{\x}\opT_{\vxi} & = & \opR_{\x-\frac{\vxi}{2}}
\exp{\left(-\frac{i}{\hbar}\x\cdot\J\vxi\right)} & ; &
\opR_{\x_1}\opR_{\x_2} & = & \opT_{2(\x_1-\x_2)}
\exp{\left(\frac{2i}{\hbar}\x_1\cdot\J\x_2\right)}~.
\end{array}
\ee
Thus, the Weyl symbol corresponding to $ \opP_{\pm}^{\X}
\oprho\opP_{\pm}^{\X}$ is
\begin{eqnarray}
\Bigl[\opP_{\pm}^{\X}\oprho\opP_{\pm}^{\X} \Bigr]\left(\x\right) & = & 
\frac{1}{2}\left( \tr{\opR_{\x}\oprho} \pm 
\tr{\opR_{\x}\opR_{\X}\oprho} \pm \tr{\opR_{\x}\oprho\opR_{\X}} +
\tr{\opR_{\x}\opR_{\X}\oprho\opR_{\X}}\right) \cr
~ & = & \frac{1}{2}\left( \tr{\opR_{\x}\oprho} + \tr{\opR_{2\X-\x}\oprho}
\pm 2 \Re{ \exp{\left( -\frac{2i}{\hbar}\x\cdot\J\X \right)} 
\tr{\opT_{2(\X-\x)}\oprho} } \right).
\end{eqnarray}
Therefore the first two terms lead directly to Wigner functions and we obtain (\ref{Wig-proj-dens}) by use of the alternative definition of the chord representation\cite{Alm98},
\be
\tA(\vxi) = \tr{ \opT_{-\vxi}\A }.
\ee

As a first example, consider the Wigner function corresponding to a pure coherent state\cite{CohDiu:book}\cite{Sch:book}, $|\Y\rangle $, with $\langle q \rangle = Q $ and $ \langle p \rangle = P$. It is well known that, this Wigner function is 
\be
W_{\Y} = \frac{1}{\pi\hbar}\exp{\left(-\frac{(\x-\Y)^2}{\hbar}\right)}
= \frac{1}{\pi\hbar}\exp{\left(-\frac{(p-P)^2+(q-Q)^2}{\hbar}\right)},
\ee
i. e. just a minimum uncertainty Gaussian. The measurement of parity with respect to the origin\cite{BruHar92} produces one of the alternatives allowed by (\ref{Wig-proj-dens}), which can be interpreted as pure Wigner functions corresponding to the sum or the difference of the coherent states $|\Y\rangle $ and $ |-\Y\rangle$:
\be
W_{\pm}^{\zero}(\x) = \frac{1}{2\pi\hbar}\frac{
\exp{\left(-\frac{(\x-\Y)^2}{\hbar}\right)} + 
\exp{\left(-\frac{(\x+\Y)^2}{\hbar}\right)} \pm
2\exp{\left(-\frac{\x^2}{\hbar}\right)}
\cos{\left(\frac{2\x\cdot\J\Y}{\hbar}\right)} }
{ 1 \pm \pi\hbar \exp{\left(-\frac{\Y^2}{\hbar}\right)} }
\ee
For sufficiently large components of $\Y $, $W_{\Y}(\zero) $ is very small, so we have nearly the same probability to obtain the state $|+\rangle $, corresponding to $W_+^{\zero}(\x) $, as the state $|-\rangle $. Both these projected Wigner functions resolve into three separate Gaussians. Those centred on $\pm\Y $ are smooth, whereas the Gaussian at the origin is modulated by fringes. These states are sometimes refered to as ``Schr\"{o}dinger cat states'' and it is easily verified that $W_{\pm}^{\zero}(\zero)=\pm(\pi\hbar)^{-1} $. Fig 1 presents the familiar form of $W_-^{\zero}(\x)$. The ``subplanckian'' scale of the fine oscillations near the origin is taken to be a sure sign of quantum coherence\cite{Zur01}.
\begin{figure}
\begin{center}
\includegraphics[width=10cm, height=6cm]{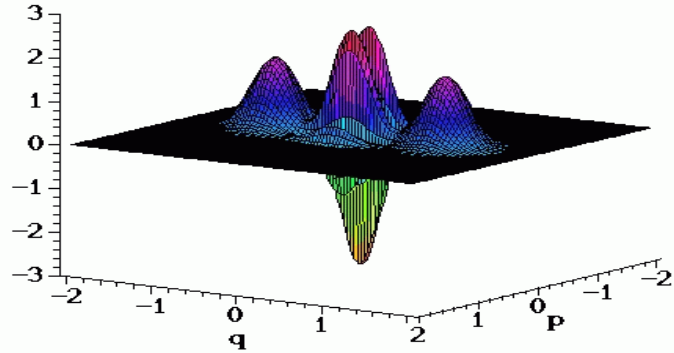}
\caption{Wigner function of an odd superposition of two coherent states in units where $\hbar=0.1$. The horizontal plane is the phase space,
$ \x=(p,q)$.}
\end{center}
\end{figure} 
Already in this simple example, we met the strangeness peculiar to parity measurements. The linearity of quantum mechanics allows us to describe the states $|+\rangle$ and $|-\rangle$ as alternative superpositions of the states $|\Y\rangle $ and $ |-\Y\rangle$. 
It may seem perverse, but we can equally describe the latter classical-like states as particular superpositions of the Schr\"{o}dinger cats, $|+\rangle $ and $|-\rangle $. Indeed, an ideal parity measurement enforces this unintuitive interpretation. Since the projections of the Wigner function provide real probabilities, it follows that, after the parity measurement, the position $-Q $ is just as likely as $Q $ and, likewise, the momentum $-P $ and $P $ are equally probable, even though the negative options would be most unprobable in the initial state. 
If the system is not completely isolated from the external environment, even an initially pure state evolves into a mixture, i.e. the density operator develops into a probability distribution over pure state densities. The main effect is to cancel the fine interference fringes characteristic of the pure quantum states, leading to more classical-like Wigner functions, as seen in Fig.2. Indeed, for an important subclass of Markovian open systems\cite{GiuJoo:book}\cite{Per:book} (random environment with no memory) the Wigner function becomes positive at a definite time whatever the initial pure state\cite{DioKie02}\cite{BroAlm03}.
\begin{figure}
\begin{center}
\includegraphics[width=10cm, height=6cm]{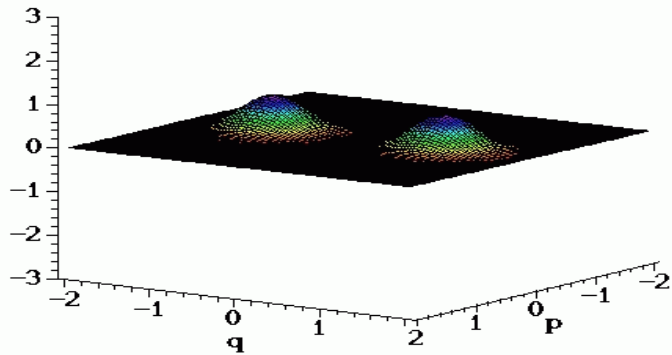}
\caption{Decoherent evolution of the Wigner function in Fig. 1 at the positivity threshold.}
\end{center}
\end{figure}
A simple example is a system in which we neglect the action of an internal Hamiltonian while allowing for linear coupling with the environment through the position $\oq $ and the momentum $ \op$. The Fokker-Planck equation\cite{GiuJoo:book}\cite{BroAlm03} that determines the evolution of the Wigner function reduces to
\be
\frac{d}{dt}W(\x,t)=\frac{\hbar}{2}c^2\left(
\frac{\der^2 W}{\der p^2}(\x,t) + \frac{\der^2 W}{\der q^2}(\x,t)
\right),
\ee
where $ c$ is the coupling constant and the solution is                        \be
W(\x,t)=\frac{1}{2\pi\hbar}\int~d\y~W(\y-\x,0)
\exp{\left( - \frac{\y^2}{2\hbar c^2 t} \right) }.
\label{evol-wig}
\ee                                                                            

This effect of the environment that progressively coarse-grains an initial pure state is more general than would appear in our simple model. Internal motion and dissipative coupling to the environment can also be included\cite{BroAlm03}. Proceeding, though, with the evolution (\ref{evol-wig}) for the initial Schr\"{o}dinger cat state $W(\x,0)=W_-^{\zero}(\x) $, we obtain
\begin{eqnarray}
W(\x,t) & = & \frac{N}{\pi\hbar(2c^2 t+1)}\Bigl[
\exp{\left(-\frac{(\x-\Y)^2}{\hbar(2c^2 t+1)}\right)} + 
\exp{\left(-\frac{(\x+\Y)^2}{\hbar(2c^2 t+1)}\right)} \cr
~ & ~ & - 2\exp{\left(-\frac{\x^2}{\hbar(2c^2 t+1)}\right)}
\exp{\left(-\frac{2c^2 t\Y^2}{\hbar(2c^2 t+1)}\right)}
\cos{\left(\frac{2\x\cdot\J\Y}{\hbar(2c^2 t+1)}\right)} \Bigr]
\end{eqnarray}
with $N^{-1}=2\left( 1 - exp{(-\Y^2/\hbar)} \right) $. Thus, the positivity threshold is $t_0=1/(2c^2) $ in this case. The full Wigner function $W(\x,t_0) $ is shown in Fig. 2. One should be aware that the symmetry of $ W(\x,t_0)$ as regards to $\zero $ has nothing to do with the parity of the mixture of states it represents. In fact, one has $W(\zero,t_0)=0 $, which shows that the probabilities of an even or odd parity measurement are actually equal. 

The result (\ref{Wig-proj-dens}) of a further ideal odd parity measurement on the mixed Wigner function $W(\x,t_0)=0 $ is displayed in Fig. 3. Again the value of the Wigner function is brought down to its minimum $-(\pi\hbar)^{-1} $, but the neighbouring interference fringes are only partly regenerated by the measurement. Thus, the hybrid nature of the state, which is pure only as concerns parity with respect to $\opR_{\zero} $, is graphically exhibited by its Wigner function.
\begin{figure}
\begin{center}
\includegraphics[width=10cm, height=6cm]{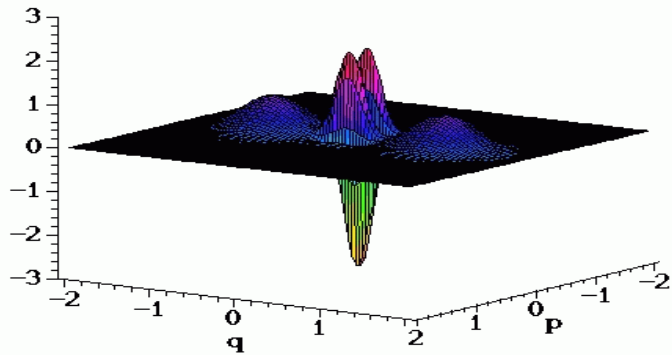}
\caption{Wigner function after an odd measurement carried out on the mixture represented in Fig.2.}
\end{center}
\end{figure}	
Allowing this spiky state to interact with the environment as before, we immediately verify that the corresponding Wigner function  becomes positive again as soon as the further interval $t_0 $ has passed, just as if it were a pure state. This follows from a simple extension of our previous arguments\cite{BroAlm03}. Ordinary mixed states loose the negative regions of their Wigner function before pure states, but odd parity mixtures must await for the pure state threshold. This depends only on the parameters of the internal quadratic Hamiltonian and of the linear coupling to the environment. 
If the initial state $|-\rangle $ evolves for a longer time in contact with the outside environment, the two mounds in Fig.2 erode even further and eventually interpenetrate. Fig. 4 shows a profile of the sharp spike that is superimposed on this smooth classical background by a positive parity measurement. Note the small negative ripples, which are tell-tails of quantum coherence.
So far, all our computations support the conjecture that $W_+(\x)$, as well as $W_-(\x)$, always take on negative values, no matter how far decoherence has proceeded prior to the parity measurement.
\begin{figure}
\begin{center}
\includegraphics[width=10cm, height=6cm]{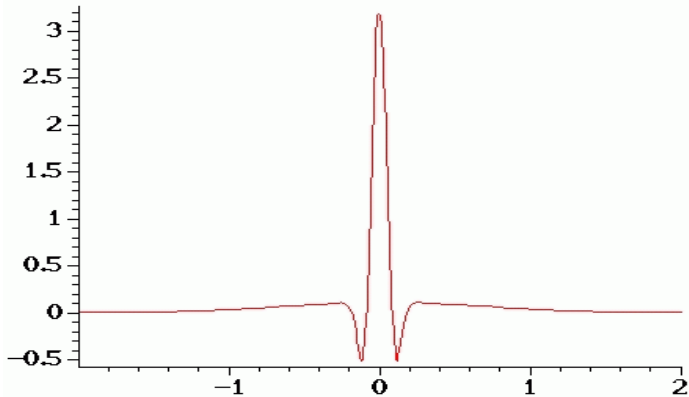}
\caption{Profile along a diagonal direction in phase space of the Wigner function reduced by an even measurement far beyond the positivity threshold.}
\end{center}
\end{figure}

	The quantum strangeness of parity reduced mixed states, that is so strikingly revealed by their Wigner function, arises from the contrast with probabilities for classical particles. However, classical waves and their Wigner functions\cite{BarBre80} are another matter. Any well tuned ensemble of clarinets is capable of producing sound waves where the odd harmonics of the fundamental note are missing. It should be pointed out that there is no relation of such classical standing waves and their harmonics, with the odd or even number of quantized photons in an optical cavity, which are all of the same frequency. All the same, there is a sense in which the manipulation of ideal parity measurements imposes the waviness of quantum matter. For example, if the parity of a mixed state of photons in a cavity is measured and immediately afterwards a photon escapes and is detected, the main effect should be the reversal of the sign of the central spike\cite{EngMor02}. To what extent real laboratory experiments will be able to evince the full features of spiky Wigner functions remains to be seen. The initial experiments in quantum optics involving single atom masers\cite{BerAuf02} are impressive, but, so far, they have been dedicated to the measurement of the Wigner function, rather than to the production of a new kind of quantum state. 

All the formulae in this paper have been presented for systems with a single degree of freedom, but they are easily generalized by extending the matrix $\J $ to higher dimensions and by suitably altering the powers of $2\pi\hbar $.

\section*{Acknowledgements}
 We thank Luiz Davidovitch, Ruynet Matos Filho and Fabricio Toscano for enlightening conversations about quantum optics and Caio H. Lewnkopf  and Raul O. Vallejos for critical comments on the manuscript. We acknowledge financial support from Faperj, CNPq, Pronex and Instituto do Milênio de Informação Quantica.

\end{document}